\documentclass{rspublic}
\usepackage{graphicx}
\input{notebook.sty}
\begin{document}
\title[Instability of solitary waves on Euler's elastica]
{Instability of solitary waves on Euler's elastica
}

\author[A. Il'ichev]{Andrej T. Il'ichev}

\affiliation{Steklov Mathematical Insitute, 119991 Moscow, Gubkina str. 8,
Russia, e-mail: ilichev@mi.ras.ru}
\label{firstpage}

\maketitle

\begin{abstract}{Euler's elastica, solitary wave, orbital stability,
unstable eigenfunction, Evans function\\
}
\\
Stability of  solitary waves  in a thin inextensible and unshearable
 rod of
infinite length
is studied. Solitary-wave  profile of
the elastica of such a
 rod without torsion has the form of a planar loop
and
its  speed depends on  a tension in the rod.
The linear instability
of a solitary-wave profile
subject
to
perturbations escaping from the plane of the loop is established for
a certain range of  solitary-wave speeds. It is done using the
properties of  the
Evans function, an analytic function
on the right complex half-plane,
that
has zeroes  if and only if there exist the unstable modes of the
linearization around a solitary-wave solution.
The result follows
from comparison of the behaviour  of the Evans function in some neighbourhood
of the origin with its asymptotic at infinity. The explicit computation
of the  leading coefficient of the Taylor series  of the Evans function
near the origin
is performed by means of the symbolic computer language.
\end{abstract}

\section{Introduction}
Dynamics of   inextensible and unshearable
thin rods of the infinite length described
in the framework of the
Bernoulli-Euler beam
model
is governed by the following equations (in dimensionless form)
\begin{eqnarray}
\label{basicsys}
& &\tau^i_{tt}=(p\tau^i)_{\xi\xi}+\tau^i_{\xi\xi}-\tau^i_{\xi\xi\xi\xi},\nonumber\\
& & \tau^i\tau_i=1,\nonumber\\
& &\tau_1\to 1, \quad\tau_{2,3}\to 0, \quad\xi\to\pm\infty,
\end{eqnarray}
where $\xi$ is the dimensionless arc-length along the elastica,
$\tau_i=\tau^i$, $i=1,2,3$ are the components of the unit vector tangent to
the elastica and $p=P-1$, $P$ being  the absolute value of  tension in
the rod; summation is assumed
under repeating indices and subscripts $t$ and $\xi$ denote differentiation
with respect to these variables.
Hereafter, the different notations for the
same components are given both with lower as well as upper indices; it is done
 for the sake of
convenience (when a component  already has the subscripts denoting
differentiation or some other) and also in the case when the rule about
summation under repeating indices is assumed.

The classification of the forms of elastica was for the first
time presented by Euler, who  derived the ordinary
differential equation of elastica (Love  [19]).
For the elastica of the infinite length
its profile  admits the configuration described by the
solitary-wave solution of (\ref{basicsys}), having the form of the
planar loop (fig.1):
\begin{eqnarray}
\label{sol}
& &\hskip-0.6cmp=-p^0=-6(1-c^2)\, {\rm sech}^{2}\sqrt{1-c^2}\zeta,\quad
\tau_3=0,
\quad
\tau_1=\tau_1^0=1-2\,{\rm sech}^{2}\sqrt{1-c^2}\zeta,\nonumber\\
& &\hskip-0.6cm\tau_2=\tau_2^0=-2\,{\rm sech}^{2}
\sqrt{1-c^2}\zeta\,{\rm sinh}\sqrt{1-c^2}\zeta,\quad
\zeta=\xi-ct,\quad 0\leq c<1.
\end{eqnarray}
The formula (\ref{sol})
 gives the particular localized solution, where the tension $P$
is distributed in the rod in a following way:
it
increases from the minimal value at the top of the loop
(for $1-c^2>1/6$,
 for example,  this value is negative, that corresponds to the compression
of the rod
at the place of the loop's localization)
to the maximal value at both infinities.
\begin{figure}
\hskip3cm\includegraphics[width=56mm,height=45mm]{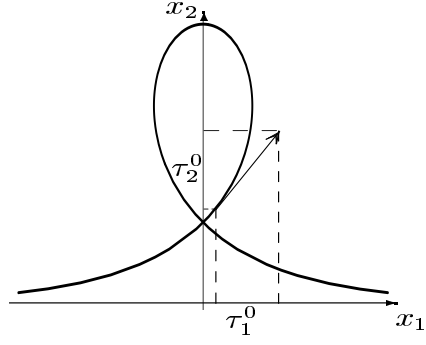}
\caption{Solitary-wave profile on an elastica and tangent vector
$\{\tau^0_1,\tau^0_2,0\}^T$.}
\end{figure}

The system of  equations (\ref{basicsys}) is a hamiltonian
one and it can be written as
\begin{eqnarray*}
\tau^i_t=\frac{\partial}{\partial\xi} \frac{\delta {H}}{\delta v_i},\quad
 v^i_t=\frac{\partial}{\partial\xi} \frac{\delta {H}}{\delta \tau_i},
\quad
\tau^i\tau_i=1,
\end{eqnarray*}
where
$${H}=\int\limits_{-\infty}^{\infty}\bigr[(v_iv^i+\tau_{i\xi}\tau^i_{\xi}+
P(\tau_i\tau^i-1)\bigl]\,d\xi.$$
Along with the hamiltonian
${H}$
one has two more conserved quantities:
$${Q}=\int\limits_{-\infty}^{\infty} (\tau^i-\tau^i_\infty)v_i\,d\xi,\quad
{R}=\int\limits_{-\infty}^{\infty} (\tau^1-1)\,d\xi,
\quad \tau^1_\infty=1,\quad \tau^{2,3}_\infty=0.
$$

In this paper we prove
linear instability of certain
solitary waves with respect to transverse perturbations
 $\delta\tau_3\neq 0$. It is shown that there exists exponentially
 growing with
time eigenfunction being the solution of the linearization about a
 solitary  wave when its dimensionless
speed does not exceed $1/\sqrt{2}$.   The existence of the unstable
eigenfunction is expected to imply the physical instability, i.\,e. the decay
of the loop under the transverse perturbations. The opposite result about the
stability, when the loop is not destroyed, takes place for
plane perturbations $\delta\tau_3=0$, $\delta\tau_{1,2}\neq 0$. Mathematically
this result is expressed by the Theorem 1.1 below.

Let us denote $\phi_c=\{\tau_1^0-1,\tau_2^0,0,v_1^0,v_2^0,0\}^T$
($v_1^0=-c(\tau_1^0-1)$, $v_2^0=-c\tau_2^0$), ${\bf w}(t)=~\{\tau_1-1,\tau_2,\tau_3,v_1,v_2,v_3\}^T$
and let
$X=H^1({\Bbb R})\times H^1({\Bbb R})\times H^1({\Bbb R})
\times L^2({\Bbb R})\times L^2({\Bbb R})\times L^2({\Bbb R})$ be
 a Hilbert space,
 $||\cdot||$ is the norm in $X$.
In
Beliaev and Il'ichev [6] using the ideas from  Grillakis {\em
et al.} [12]
the following theorem is proved
relating to the nonlinear stability
of solitary waves
(\ref{sol})
subject to perturbations
of $\tau_{1,2}^0$ only.
\begin{theorem}
\label{th}
Let   $\tau_3\equiv 0$ and
for any given ${\bf w}_0\in X$ near
$\phi_c$ in $X$, $||{\bf w}_0-\phi_c||<\gamma$, there exist
$T=T(\gamma)>0$ and a vector function ${\bf w}(t)\in C([0,T),X)$
(continuous  with values in  $X$)
such that ${\bf w}(0)={\bf w}_0=\{\tau_0^{l},0,v_0^l,0\}^T$, $l=1,2$,
and for all
$0\leq t\leq T$, $\tau_0^i\tau_{0i}=\tau_i\tau^i=1$ and
${H}({\bf w})={H}({\bf w}_0)$,
${Q}({\bf w})={Q}({\bf w}_0)$,
${R}({\bf w})={R}({\bf w}_0)$. Then, for all
$\varepsilon>0$ there exists $\delta>0$ such that
if
 $||{\bf w}_0-\phi_c||<\delta$, then
$$\sup\limits_{t>0}\inf\limits_{\omega\in{\Bbb R}}||{\bf w}(t)-\phi_c(\cdot+\omega)||
<\varepsilon.$$
\end{theorem}
The theorem was proved by demonstrating that
the set of translates of a solitary wave gives a local minimum to
the invariant functional
$${F}({\bf w})={H}({\bf w})+c Q({\bf w})-{R}({\bf w})
$$
on  the closed submanifold $\tau_i\tau^i=1$ in $X$.
This, in its turn, follows from the fact that the `linearized
hamiltonian' ${\cal H}(\phi_c)=\delta^2\,F(\phi_c)/\delta^2{\bf w}$ has exactly one zero
eigenvalue and positive spectrum bounded away from zero.
The analysis of Beliaev and Il'ichev [6] is extended in
Dichmann {\it et al.} [9]
to the case
when the rotational kinetic energy
(which is small compared to the total energy for thin rods)
is not neglected
 and the same stability result for
subsonic solitary waves was established.
For $\delta\tau_3=\linebreak =\tau\neq 0$, however,
the operator ${\cal H}(\phi_c)$ has an extra negative eigenvalue and
the sufficient conditions
for the stability of bound states as they appear in
Grillakis {\it et al.} [12]
 are violated.
The system of equations (\ref{basicsys}) is also invariant under group
of rotations (see sec. 6). However,
the stability theory  given in
Grillakis {\it et al.}
[13] for hamiltonian systems with several symmetries
is not applicable in the case under consideration by the reason
 that solitary wave orbit is in fact the
set of translates of a solitary wave, and therefore
the conserved quantity associated with the rotational symmetry, being nonlocal,
identically equals zero at a solitary wave.

The
linearized
equations (\ref{basicsys}) for $\delta\tau_{1,2}$ and $\tau$  decouple
and we analyze here only the equation for $\tau$ since the stability result
in the planar case is known.
A solitary wave is called linearly unstable
if there exist
solutions of the linearized equation  of the form
\begin{eqnarray}
\label{vid}
\tau=\tau(\xi,t)={\rm e}^{\lambda t} w(\lambda,\zeta)
\end{eqnarray}
with  $w(\lambda,\zeta)$ decaying exponentially as $\zeta\to\pm\infty$ and
$\Real{\lambda}>0$.
The  function $w(\lambda,\zeta)$ obeys the eigenvalue problem
\begin{eqnarray}
\label{basicsolu}
 (\lambda-c \frac{d}{d\zeta})^2w= \frac{d^2}{d\zeta^2}w-
 \frac{d^4}{d\zeta^4}w-\frac{d^2}{d\zeta^2}(p^0 w).
\end{eqnarray}
The
equation (\ref{basicsolu}) has the same form as the corresponding equation
for the stability problem  discussed in
Alexander and Sachs [2].
That is why, the theoretical results  of this paper are also valid
for our
analysis.
Yet, the  coefficient $p^0$ in
(\ref{basicsolu}) does not coincide  as a function of $\zeta$
with the corresponding coefficient in
Alexander and Sachs [2]
and, as a consequence,  our computations and the result
are different from the computations and the result of these authors.

We get the instability result by means
of the Evans function.
The Evans function $D(\lambda)$ is
constructed as
an analytic
on the right complex half-plane
function of the
spectral  parameter;
$D(\lambda)$
has zeroes  if and only if there exist the unstable modes of the
linearization around a solitary-wave solution.
The use of an Evans function to get instability result for solitary waves
appears in Pego and Weinstein [20]. In parabolic problems,
the ideas of
Evans [10] were further developed by Jones [15]
and Alexander {\it et
al.} [3] (see also Kapitula [16] and references therein).
Alexander and Sachs [2]  extended
 the technique  of Pego and Weinstein [20] to the case
of the Bousssinesq-type equations where two decaying modes  are present in
the model.
The application of the Evans function method for instability problems
in various fields of hydrodynamics and physics can be found in
Swinton and Eglin [22],
Pego {\it et
al.} [21],
Alexander {\it et al.} [4], Gardner and Zumburin [11],
Kapitula [17], Kapitula and Sandstede [18],
Afendikov and Bridges [1], Bridges {\it et al.}  [8].

The Evans function is real on the real axis and tends to unity  as
$|\lambda|\to\infty$.
Our instability result follows
from comparison of the behaviour  of the Evans function in some neighbourhood
of the origin with its asymptotic at infinity. We show that in a small
neighbourhood of the origin one has $D(\lambda)<0$ for real positive $\lambda$.
More precisely, the following theorem holds.
\begin{theorem}
The Evans function $D(\lambda)$ constructed for the eigenfunction
$w(\lambda,\zeta)$
from {\rm (\ref{vid})} is analytic in a neighbourhood of the origin
and its converging Taylor series in this neighbourhood reads
\begin{eqnarray}
\label{DEvans}
D(\lambda)=
-\frac{1-2c^2}{4(1-c^2)^3}\lambda^2+\sum\limits_{n=3}^\infty e_n(c)\lambda^n.
\end{eqnarray}
\end{theorem}
It follows then, that
for $c^2<1/2$
the  function $D(\lambda)$ has a zero somewhere at $\lambda=\lambda_0$ on the
positive real axis, that, in its turn, implies the existence of the
unstable eigenfunction $w(\lambda_0,\zeta)$ from (\ref{vid}).

In section 2 we give the formulation and linear stability problem.
In section 3 we discuss analytic properties of the solutions of the linearized
system as $\lambda$ varies in a zero neighbourhood. In section 4 the Evans function is introduced and its behaviour for
large $|\lambda|$ is described. In section 5 we describe computations
(made by means of
 the
 symbolic language {\tt Mathematica 4.0}) of the
leading coefficient of the Taylor series of $D(\lambda)$ in a
neigbourhood of the origin.
The explicit form of key expressions in our computations
are listed in Appendix A.  In section 6 we discuss our conclusions.

\section{Formulation  and linear stability problem}
\setcounter{equation}{0}
We consider a thin inextensible and unshearable elastic rod
of infinite length initially coinciding with the
$x_1$ axis of a Cartesian coordinate system.
The total energy of such a rod is  the sum of the kinetic energy
and the bending energy, the torsion energy is neglected.
The respective energy densities
 $K$ and $W$ are given by the expressions
$$ K=\frac{1}{2}\rho S x^i_t x_{it},\quad
 W=\frac{1}{2}I E x_{\xi\xi}^i x_{i\xi\xi},
$$
where $x_i$, $i=1,2,3$
are the coordinates of the points on a neutral line (elastica) of the rod, $\rho$
is the density of the rod,
$S$ is the area of the cross section of the rod, $\rho I$
is the moment of inertia of the cross section of the rod about the line,
orthogonal to the principal plane of bending
$x_1 x_2$, $E$ is the Young module, $\xi$ is  the arc-length along the elastica.
The elastica is given by the equations
$x^i=x^i(\xi,t)$.
For thin rods the rotational part of the kinetic energy is small in comparison
with the kinetic energy of points of the elastica
[6] and we neglect it here.

The equations of motion can be obtained by taking the variations of the
lagrangian $\Lambda$,
\begin{eqnarray*}
\label{lagr}
\Lambda=\frac{1}{2}\int\limits_{t_0}^t\int\limits_{-\infty}^{\infty}
(\rho S x^i_t x_{it}-IEx^i_{\xi\xi}x_{i\xi\xi})d\xi\,dt
\end{eqnarray*}
under the condition of inextensibility
$$x_{i\xi}x^i_{\xi}=1.$$
These equations read
\begin{eqnarray}
\label{basicsys0}
& &\rho S x^i_{tt}= (P x_{i\xi})_{\xi}-IE x_{i\xi\xi\xi\xi},\quad
 x_{i\xi}x^i_{\xi}=1,
\end{eqnarray}
where
$P(\xi,t)=p(\xi,t)+p_\infty$ is the Lagrange multiplier,
$p\to 0$ as $\xi\to\pm \infty$.
Making the scaling transformations  in (\ref{basicsys0})
\begin{eqnarray*}
p\to p_\infty p, \quad\xi\to \sqrt{I E/p_\infty}\xi,\quad
t\to\sqrt{\rho S I E/p_\infty^2} t,
\end{eqnarray*}
and preserving the old notations
one gets (\ref{basicsys}) with $\tau_i=x^i_\xi$.

Linearizing the equations
(\ref{basicsys})
about the solitary-wave solution
 (\ref{sol}) we get the  following equation for the perturbation
$\delta\tau_3=\tau$
of the third component of the tangent
vector (\ref{sol})
\begin{eqnarray}
\label{vozm}
\tau_{tt}=-(p^0\tau)_{\xi\xi}+\tau_{\xi\xi}-\tau_{\xi\xi\xi\xi}.
\end{eqnarray}
We seek the solution of (\ref{vozm}) in the form
(\ref{vid}).
After  substitution  of (\ref{vid}) into the equation (\ref{vozm}) it
transforms into the equation (\ref{basicsolu}).
The last equation, in its turn, can be written in the matrix form
\begin{eqnarray}
\label{matr1}
{\bf y}'={\cal M}(\lambda,\zeta){\bf y},
\end{eqnarray}
$$ {\bf y}=\{y_1,y_2,y_3,y_4\}^T,\,\,
y_1=w,\,\, y_2=w',\,\, y_3=w'',\,\, y_4=w''',
$$
where prime denotes differentiation with respect to
 $\zeta$, and
$$
{\cal M}(\lambda,\zeta)=\left(
\begin{array}{cccc}
0 & 1& 0& 0\\ 0 & 0 &1 & 0\\
0 & 0& 0& 1\\-\lambda^2-p^{0''} & 2\lambda c-2p^{0'} & 1-c^2-p^0 &0
\end{array}
\right).
$$

The adjoint equation reads
\begin{eqnarray}
\label{sopr}
(\lambda+c\frac{d}{d\zeta})^2w^*=\frac{d^2}{d\zeta^2}w^*-
\frac{d^4}{d\zeta^4}w^*-p^0\frac{d^2}{d\zeta^2}w^*,
\end{eqnarray}
or in the matrix form
\begin{eqnarray}
\label{matr2}
{\bf z}'=-{\bf z}{\cal M}(\lambda,\zeta),
\end{eqnarray}
\begin{eqnarray*}
& &{\bf z}=\{z_1,z_2,z_3,z_4\},\,\,
z_4=w^*,\,\,z_3=-w^{*'},\,\, z_2=w^{*''}-(1-c^2-p^0)w^*,\\
& &z_1=-w^{*'''}+(1-c^2-p^0)w^{*'}-(2\lambda c-p^{0'})w^*.
\end{eqnarray*}

\section{Analytic properties of the solutions of the linearized system for
$\lambda$ in the zero neighbourhood}
In this section we enumerate the results of Alexander and Sachs [2]
concerning the analytic properties of the solutions of systems
having the form of
(\ref{matr1}), (\ref{matr2}) when $\lambda$
 locates close to
zero.

Let us denote ${\cal M}_\infty(\lambda)=
\lim\limits_{\zeta\to\pm\infty}{\cal M}(\lambda,\zeta)$,
 $\mu_\alpha(\lambda)$ ($\alpha=1,2,3,4$)
the eigenvalues of
${\cal M}_\infty(\lambda)$,
${\bf r}_\alpha(\lambda)$ the right and   ${\bf l}_\alpha(\lambda)$
the left
eigenvectors of the matrix
 ${\cal M}_\infty(\lambda)$.  The matrix ${\cal M}_\infty(\lambda)$  is
 given by the same expression as the
corresponding matrix in Alexander and Sachs [2].
For $\lambda=0$ the vectors labeled by the subscript $4$ represent
the generalized eigenvectors of ${\cal M}_\infty(\lambda)$
(see formulas (3.9), (3.10) in Alexander and
Sachs [2]).
The eigenvectors are normalized so, that the first
(for right eigenvectors) and last (for left eigenvectors)
components are unity.

The characteristic equation ${\rm det}\,[{\cal M}_\infty(\lambda)-\mu {\cal
E}]=0$,
where ${\cal E}$ denotes the unit matrix,
has the form
\begin{eqnarray}
\label{vek}
\mu^4-(1-c^2)\mu^2-2\mu c\lambda+\lambda^2=0.
\end{eqnarray}
In Alexander and Sachs [2] the following lemma is proved.
\begin{lemma}
For ${\rm Re}\,\lambda\neq 0$ the equation
{\rm (\ref{vek})}
has two roots in each complex half-plane.
\end{lemma}
\vskip0.2cm
Following  [2] we denote
$\mu_1(\lambda)$, $\mu_3(\lambda)$
the roots,
lying in the left complex half-plane for
 $ \Real\lambda>0$
($|\Real\mu_1(\lambda)|>
|\Real\mu_3(\lambda)|$ for $\lambda$ in a zero neighbourhood).
In the neighborhood of the origin
the roots $\mu_k(\lambda)$, $k=1,3$
have the asymptotic form
[2]
\begin{eqnarray}
\label{assmu}
& &\mu_1(\lambda)=-\sqrt{1-c^2}+\frac{c\lambda	}{1-c^2}+
\frac{1+2c^2}{2(1-c^2)^{5/2}}\lambda^2+O(\lambda^3),\nonumber\\
& &\mu_3(\lambda)=-\frac{\lambda}{1-c}+O(\lambda^3).
\end{eqnarray}
The eigenvalues
 $\mu_{1,3}(\lambda)$ and the eigenvectors
${\bf r}_{1,3}(\lambda)$, ${\bf l}_{1,3}(\lambda)$
may lose the analyticity property in the points where
the eigenvalues are not simple any more or, in other words, when
the corresponding roots of equation
(\ref{vek}) are  multiple. The roots of
(\ref{vek}) become multiple
 when the resultant of the
polynomial in the left hand side of (\ref{vek}) and its derivative equals
zero. The resultant is
$$16\lambda^2[(1-c^2)^3+(-8-20 c^2+c^4)\lambda^2+16\lambda^4],$$
and it equals zero when
$\lambda_{1,2}=0,$  and
\begin{eqnarray}
\label{branch}
& &\lambda_{3,4}=\pm\frac{\sqrt{8+20c^2-c^4+c(8+c^2)^{3/2}}}{4\sqrt{2}},
\nonumber\\
& &\lambda_{5,6}=\pm\frac{\sqrt{8+20c^2-c^4-c(8+c^2)^{3/2}}}{4\sqrt{2}}.
\end{eqnarray}
The nonzero branches (\ref{branch}) are pictured in Fig.2. It is seen that
for  nonzero $\lambda$ the points where the eigenvalues are multiple are
separated from zero when $0\leq c<1$. It is also seen that when $\lambda$
crosses the imaginary axis the eigenvalues $\mu_3$ and $\mu_4$, coinciding
with zero
at $\lambda=0$, escaping from their complex half-planes.
\begin{figure}
\hskip3cm\includegraphics[width=56mm,height=55mm]{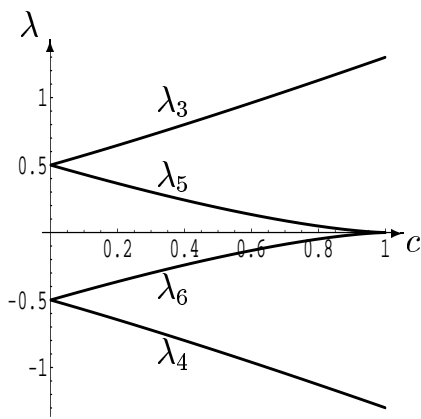}
\caption{ Curves $\lambda$ versus $c$  where the eigenvalues are multiple.
Along the branch $\lambda=\lambda_3$ the eigenvalues $\mu_2$ and $\mu_4$
coincide; along the branch $\lambda=\lambda_5$  the eigenvalues $\mu_1$ and
$\mu_3$ coincide; along the branch $\lambda=\lambda_6$ the eigenvalues $\mu_3$ and $\mu_2$
coincide and along the branch $\lambda=\lambda_4$ the eigenvalues $\mu_1$ and
$\mu_4$ coincide.}
\end{figure}

At $\lambda=0$ one has $\mu_3(0)=\mu_4(0)=0$. Yet, the kernel
of ${\cal M}_\infty(\lambda)-\mu_3(\lambda){\cal E}$
spanned by ${\bf r}_3(\lambda)$
(${\bf l}_3(\lambda)$)
is uniformly one dimensional and the vectors themselves are continuous at
$\lambda=0$ (see formulas (3.9), (3.10) in Alexander and Sachs [2]).
Hence, in a small enough neighbourhood of the origin, not including the
points given by (\ref{branch}), the normalized vectors
${\bf r}_{1,3}(\lambda)$ and
${\bf l}_{1,3}(\lambda)$ are analytic.

It follows from general theory of the ordinary differential equations that
there exist solutions
of (\ref{matr1}), (\ref{matr2}),
satisfying
\begin{eqnarray}
\label{assyz}
 & &\lim\limits_{\zeta\to\infty}{\rm e}^{-\mu_k(\lambda)}{\bf y}_k
(\lambda,\zeta)={\bf r}_k(\lambda),\nonumber\\
& &\lim\limits_{\zeta\to-\infty}{\rm e}^{\mu_k(\lambda)}{\bf z}_k
(\lambda,\zeta)={\bf l}_k(\lambda),\quad k=1,3.
\end{eqnarray}

Analyticity of
${\bf r}_{1,3}(\lambda)$
${\bf l}_{1,3}(\lambda)$  implies the following theorem, proved in
Pego and Weinstein [20] and generalized for the two mode case in
Alexander and Sachs [2].

\begin{theorem}
 Solutions
${\bf y}_{1,3}(\lambda,\zeta)$,
${\bf z}_{1,3}(\lambda,\zeta)$ are analytic in some
neighborhood of $\lambda=0$.
\end{theorem}
\section{Evans function}
\setcounter{equation}{0}
Consider the vector fields
${ \bf y}^{\wedge}(\lambda,\zeta)$, ${ \bf z}^{\wedge}(\lambda,\zeta)$
with the components
\begin{eqnarray}
\label{extform}
y^{\wedge}_{\alpha\wedge \beta}=
y_{1\alpha}y_{3\beta}-y_{1\beta}y_{3\alpha},\quad
z^{\wedge}_{\alpha\wedge \beta}= z_{1\alpha}z_{3\beta}-z_{1\beta}z_{3\alpha},
\hskip0.2cm\alpha>\beta,\hskip0.2cm \alpha,\beta=1,2,3,4,
\end{eqnarray}
where $y_{k\alpha}$, $z_{k\alpha}$ are the components of the vectors
${\bf y}_k$, ${\bf z_k}$, correspondingly.
We number the pairs $\alpha\wedge \beta$ by the following way:
$1\wedge2\to 1$, $1\wedge 3\to 2$, $1\wedge4\to 3$, $2\wedge 3\to 4$,
$2\wedge 4\to 5$, $3\wedge 4\to 6$.
The vectors ${\bf y}^\wedge(\lambda,\zeta)$,
${\bf z}^\wedge(\lambda,\zeta)$ obey the equations
\begin{eqnarray}
\label{eqnwedge}
{\bf y}^{\wedge'}={\cal M}^{\wedge}(\lambda,\zeta){\bf y}^{\wedge},\quad
{\bf z}^{\wedge'}=-{\bf z}^{\wedge}
{\cal M}^{\wedge}(\lambda,\zeta),
\end{eqnarray}
where
$$
{\cal M}^\wedge(\lambda,\zeta)=\left(
\begin{array}{cccccc}
0&1&0&0&0&0\\ 0&0&1&1&0&0\\
2\lambda c-2p^{0'}&1-c^2-p^0&0&0&1&0\\ 0&0&0&0&1&0\\
\lambda^2+p^{0''}&0&0&1-c^2-p^0&0&1\\
0&\lambda^2+p^{0''}&0&-2\lambda c+2p^{0'}&0&0
\end{array}
\right).
$$
The eigenvalues of
 ${\cal M}^{\wedge}_\infty(\lambda)={\cal M}^{\wedge}(\lambda,\pm\infty)$
are given by
$\mu_\alpha(\lambda)
+\mu_\beta(\lambda)$.
The eigenvalue
 $\mu^\wedge(\lambda)=\mu_1(\lambda)+\mu_3(\lambda)$  with the minimal real part
is simple
in the right complex half-plane
 ${\rm Re}\,\lambda>0$. Therefore, the  right
${\bf r}^\wedge(\lambda)$  and left
${\bf l}^\wedge(\lambda)$ eigenvectors of the matrix
${\cal M}^\wedge_\infty(\lambda)$
associated with this eigenvalue
are analytic everywhere in the right complex
 $\lambda$-half-plane.   This, in its turn, implies the analyticity of
the vector-functions
${ \bf y}^{\wedge}(\lambda,\zeta)$ and ${ \bf z}^{\wedge}(\lambda,\zeta)$
everywhere in the right complex $\lambda$-half-plane.
The components of the vectors
${\bf r}^\wedge(\lambda)$ and
${\bf l}^\wedge(\lambda)$  are given via the components of the vectors
${\bf r}_{1,3}(\lambda)$ and
${\bf l}_{1,3}(\lambda)$ by the expressions similar to (\ref{extform})
everywhere
where the last vectors are well defined, i.\,e. in some neighbourhood of
the origin of the complex $\lambda$-plane.   The
vector-solutions ${\bf y}^\wedge$  and
${\bf z}^\wedge$ satisfy
\begin{eqnarray*}
\label{assyz1}
 & &\lim\limits_{\zeta\to\infty}{\rm e}^{-\mu^\wedge(\lambda)}{\bf y}^\wedge
(\lambda,\zeta)={\bf r}^\wedge(\lambda),\nonumber\\
& &\lim\limits_{\zeta\to-\infty}{\rm e}^{\mu^\wedge(\lambda)}{\bf z}^\wedge
(\lambda,\zeta)={\bf l}^\wedge(\lambda).
\end{eqnarray*}

The
analytic in the complex half-plane ${\rm Re}\,\lambda>0$
Evans function
$\hat{D}(\lambda)$
is defined as follows:
\begin{eqnarray}
\label{Evans}
\hat{D}(\lambda)=
{\bf z}^{\wedge}\cdot{\bf y}^{\wedge}=
\det\left(
\begin{array}{cc}
{\bf z}_1(\lambda,\zeta)\cdot
{\bf y}_1(\lambda,\zeta)&
{\bf z}_1(\lambda,\zeta)\cdot{\bf y}_3(\lambda,\zeta)\\
{\bf z}_3(\lambda,\zeta)\cdot{\bf y}_1(\lambda,\zeta)&
{\bf z}_3(\lambda,\zeta)\cdot{\bf y}_3(\lambda,\zeta)
\end{array}
\right).
\end{eqnarray}
The last equality in (\ref{Evans})
has the sense  where the determinant in the right hand
side is well defined, in particular, in the neighborhood of the origin, where,
according to Theorem 3.2, the vector-functions
${\bf y}_k(\lambda,\zeta)$, ${\bf z}_k(\lambda,\zeta)$, $k=1,3$
are analytic.
From (\ref{eqnwedge}) it follows that $\hat{D}(\lambda)$ is independent on $\zeta$.
It also follows that the function
 $\hat{D}(\lambda)$ is analytic in the neighborhood of the origin
of the complex $\lambda$-plane.
\begin{theorem}
 {\rm ([2], [3])}. For ${\rm Re}\,\lambda>0$
the function
 $\hat{D}(\lambda)$ has zeroes if and only if there is a solution of
 {\rm  (\ref{basicsolu})},
which decays exponentially as $\zeta\to\pm\infty$.
\end{theorem}

The asymptotic behaviour of the Evans function $\hat{D}(\lambda)$
as $\lambda\to\infty$ is governed
by the following
\begin{theorem}
$\hat{D}(\lambda)\to {\bf l}^\wedge(\lambda)\cdot {\bf r}^\wedge(\lambda)$
as $|\lambda|\to \infty$.
\end{theorem}

To prove Theorem 4.2 it is sufficient to verify the conditions of
proposition  1.17 in Pego and Weinstein [20]
(we note that in [20]  the vectors corresponding to
${\bf r}^\wedge(\lambda)$ and ${\bf l}^{\wedge}(\lambda)$
are normalized so that
${\bf l}^\wedge(\lambda)\cdot
{\bf r}^{\wedge}(\lambda)=1$, and their Evans function has asymptotic value
unity).
It has to be checked
that
\\
$\bullet$ ${\cal M}^{\wedge}_\infty(\lambda)$ is diagonalizable for large
$\lambda$;  \\
$\bullet$   ${\cal F}(\lambda,\zeta)={\cal W}(\lambda){\cal R}(\zeta)
{\cal V}(\lambda)\to 0$ as $|\lambda|\to\infty$,
where ${\cal V}(\lambda)$ is a matrix of right eigenvectors,
${\cal W}(\lambda)$ is its inverse and
${\cal R}(\zeta)={\cal M}^\wedge(\lambda,\zeta)-
{\cal M}^\wedge_\infty(\lambda)$.

The validity of the first  condition is  given by the
\begin{lemma}
{\rm ([2])}. For large $|\lambda|$ the matrix
${\cal M}^{\wedge}_\infty(\lambda)$ has eigenvalues
\begin{eqnarray}
\label{asseigen}
\pm\sqrt{2\lambda}\bigl(1+O(|\lambda|^{-1})\bigr),
\quad
\pm\ri\sqrt{2\lambda}\bigl(1+O(|\lambda|^{-1})\bigr),
\quad \pm\ri c\bigl(1+O(|\lambda|^{-2})\bigr).
\end{eqnarray}
\end{lemma}
The second condition is verified by direct computation using
{\tt Mathematica 4.0}.
Consider the eigenvectors ${\bf v}_i^\pm$, $i=1,2,3$  associated with the
eigenvalues (\ref{asseigen}) with first component normalized to unity.
One has ${\bf v}_i^\pm=\hat{\bf v}_i^\pm \bigr(1+O(|\lambda|^{-1})\bigl)$, where
\begin{eqnarray*}
& & \hat{\bf v}_1^\pm=
\left(1;\pm \sqrt{2\lambda};\frac{1-c^2\pm c\sqrt{2\lambda}+2\lambda}{2};
\frac{-1+c^2\mp c\sqrt{2\lambda}+2\lambda}{2};\right.\\
& &\left.\mp\frac{(1-c^2\pm c\sqrt{2\lambda}-2\lambda)\sqrt{\lambda}}{\sqrt{2}};
\lambda^2\mp\frac{c\sqrt{\lambda}(-1+c^2\mp\sqrt{2\lambda}c+2\lambda)}{\sqrt{2}}
\right)^T,\\
& & \hat{\bf v}_2^\pm=
\left(1;\pm \ri\sqrt{2\lambda};\frac{1-c^2\mp \ri c\sqrt{2\lambda}-2\lambda}{2};
\frac{-1+c^2\pm \ri c\sqrt{2\lambda}-2\lambda}{2};\right.\\
& &\left.\pm\frac{\ri (-1+c^2\pm \ri c\sqrt{2\lambda}-2\lambda)\sqrt{\lambda}}{\sqrt{2}};
\lambda^2\pm\frac{\ri c\sqrt{\lambda}(-1+c^2\pm\ri \sqrt{2\lambda}c-
2\lambda)}{\sqrt{2}}
\right)^T,\\
& & \hat{\bf v}_3^\pm=
\left(1;\pm\ri c;\frac{1}{2}-c^2\mp \ri\lambda;
\frac{-1\pm 2\ri \lambda}{2};\frac{-c(\pm \ri +2\lambda)}{\sqrt{2}};
-\lambda(\pm\ri+\lambda)\right)^T.
\end{eqnarray*}
Then ${\cal V}=\hat{\cal V}\bigl(1+O(|\lambda|^{-1})\bigr)$, where
$\hat{\cal V}$ is a column matrix combined from the vectors
$\hat{\bf v}_i^\pm$. The determinant $\Delta$ of ${\cal V}$ is given by the
expression $\Delta=\hat{\Delta}\bigl(1+O(|\lambda|^{-1})\bigr)$,
where
$$\hat{\Delta}=-16\lambda^2(c^4-4\lambda^2)^2
$$
is the determinant of
$\hat{\cal V}$. Hence, both ${\cal V}$ and $\hat{\cal V}$ are invertible
for large $|\lambda|$ and ${\cal F}=\hat{\cal F}
\bigl(1+O(|\lambda|^{-1})\bigr)$, where
$\hat{\cal F}=\hat{\cal W}{\cal R}\hat{\cal V}$ and $\hat{\cal W}=\hat{\cal
V}^{-1}$. The direct computation shows that
$\hat{\cal F}=\hat{\Delta}^{-1}{\cal N}$,
where the components of the matrix ${\cal N}(\lambda,\zeta)$ are
polynomials in
$\sqrt{\lambda}$ and the maximal degree of these polynomials  is $11$.
Therefore, ${\cal F}(\lambda,\zeta)\to 0$ as $|\lambda|\to\infty$ and the
proposition 1.17 of Pego and Weinstein [20] may be used in our case.
\section{Computation of the leading coefficient of the Taylor series of
${D}(\lambda)$}
\setcounter{equation}{0}
Let us denote $Y_k$ the first, and $Z_k$  the last component
of the vectors
${\bf y}_k$ and ${\bf z}_k$, obeying at the same time
(\ref{basicsolu}), (\ref{sopr}), respectively. From (\ref{assyz}) and
(\ref{assmu}) one has
\begin{eqnarray}
\label{Y}
&& Y_1={\rm e}^{-\sqrt{1-c^2}\zeta}\left(1+\frac{c\zeta}{1-c^2}\lambda+
\left[\frac{c^2\zeta^2}{2(1-c^2)^2}+\frac{1+2c^2}{2(1-c^2)^{5/2}}
\zeta\right]\lambda^2\right)+
O(\lambda^3),\nonumber\\
& &Y_3(\lambda,\zeta)=1-\frac{\zeta}{1-c}\lambda+
\frac{1}{2}\frac{\zeta^2}{(1-c)^2}
\lambda^2+O(\lambda^3);
\end{eqnarray}
as $\zeta\to \infty$, and
\begin{eqnarray}
\label{Z}
& & Z_1={\rm e}^{\sqrt{1-c^2}\zeta}\left(1-\frac{c\zeta}{1-c^2}\lambda+
\left[\frac{c^2\zeta^2}{2(1-c^2)^2}-\frac{1+2c^2}{2(1-c^2)^{5/2}}\zeta\right]\lambda^2
\right)+
O(\lambda^3),\nonumber\\
& &Z_3(\lambda,\zeta)=1+\frac{\zeta}{1-c}\lambda+
\frac{1}{2}\frac{\zeta^2}{(1-c)^2}
\lambda^2+O(\lambda^3),
\end{eqnarray}
as $\zeta\to-\infty$.

According to Theorem 3.2 we look for solutions
of the equations (\ref{basicsolu}), ({\ref{sopr}) in the neighbourhood of
 $\lambda=0$ in the form of expansions
\begin{eqnarray*}
& &Y_k(\lambda)=Y_{k0}+\lambda Y_{k1}+\frac{1}{2}\lambda^2Y_{k2}+O(\lambda^3),
\\
& &Z_k(\lambda)=Z_{k0}+\lambda Z_{k1}+\frac{1}{2}\lambda^2 Z_{k2}+O(\lambda^3).
\end{eqnarray*}
The coefficients of the above expansions
obey the equations
\begin{eqnarray}
\label{basiceq1Y}
{\cal L}\,Y_{k0}=0,
\end{eqnarray}
\begin{eqnarray}
\label{basiceq1Z}
{\cal L}^*\,Z_{k0}=0,
\end{eqnarray}
\begin{eqnarray}
\label{basiceq2Y}
{\cal L}\,Y_{k1}=2cY_{k0}',
\end{eqnarray}
\begin{eqnarray}
\label{basiceq2Z}
\quad  {\cal L}^*\,Z_{k1}=-2c Z_{k0}',
\end{eqnarray}
\begin{eqnarray}
\label{basiceq3Y}
{\cal L}\,Y_{k2}=-2 Y_{k0}+4c Y_{k1}',
\end{eqnarray}
\begin{eqnarray}
\label{basiceq3Z}
{\cal L}^*\,Z_{k2}=-2Z_{k0}-4cZ_{k1}',
\end{eqnarray}
where the differential operators
 ${\cal L}$ and ${\cal L}^*$ are defined as follows:
$${\cal L}Y=Y''''-(1-c^2)Y''+(p^0 Y)'',\quad
{\cal L}^*Z=Z''''-(1-c^2)Z''+p^0Z''.
$$
The equations
 (\ref{basiceq1Y}-\ref{basiceq3Z}) have to be solved
taking into account the conditions
(\ref{Y}), (\ref{Z}).

For $\lambda=0$ the complete bases  of
solutions of the equations (\ref{basicsolu}) and ({\ref{sopr}), which coincide
with the  equations  (\ref{basiceq1Y}) and (\ref{basiceq1Z}), are
determined explicitly.  Putting $\lambda=0$ in
(\ref{basicsolu}) and
integrating it twice we get
\begin{eqnarray}
\label{2int}
{\cal L}_0 w=\frac{d^2 w}{d\zeta^2}-(1-c^2) w+ p^0 w=a+b\zeta.
\end{eqnarray}
The solving of the equation  (\ref{basiceq1Y}) is equivalent to solving
of the second order equation (\ref{2int}) with arbitrary constants $a$ and
$b$.
When $a=b=0$ the particular
solutions are
$$w_1=\tau_2^0, \quad
w_2=w_1\int w_1^{-2}\,d\zeta.
$$
For some $a\neq0$, $b=0$ the solution of (\ref{2int}) is given by
$$w_3=-w_1\int w_2\,d\zeta+w_2\int w_1\,d\zeta.$$
For $a=0$  and some $b\neq 0$ the solution of
(\ref{2int}) is
$$w_4=-w_1\int \zeta w_2\,d\zeta+w_2\int \zeta w_1\,d\zeta.$$

The solutions $w_1$, $w_2$, $w_3$, $w_4$ constitute the basis in the
space of solutions
of the  equation  (\ref{basiceq1Y}). By the appropriate normalization
of $w_1$ and $w_3$ one gets $Y_{10}$ and $Y_{30}$ correspondingly, such that
according to  (\ref{Y}) $Y_{10}\to \re^{-s\zeta}$ and
$Y_{30}\to 1$ as $\zeta\to\infty$:
\begin{eqnarray*}
Y_{10}= \frac{\re^{s\zeta}-\re^{-s\zeta}}{(\re^{s\zeta}+\re^{-s\zeta})^2},
\quad Y_{30}=\frac{1-6\re^{2s\zeta}+\re^{4s \zeta}}{(1+\re^{2 s\zeta})^2},
\end{eqnarray*}
where $s=\sqrt{1-c^2}$ hereafter.

Normalizing $w_2$ one gets $Y_{20}$, having asymptotic
$Y_{20}\to \re^{s\zeta}$ as $\zeta\to\infty$. Linear combination of
$w_4$, $Y_{03}$ and $Y_{20}$ gives $Y_{40}$ asymptotic to $\zeta$ as
$\zeta\to\infty$.

Putting $\lambda=0$ in (\ref{sopr}) we get the  equation
(\ref{basiceq1Z}). Following  [2]  we
denote $w^*{''}=\phi$. Then $\phi=w_1$
and $\phi=w_2$ are two solutions of
(\ref{2int}) with $a=b=0$. Therefore,
$$w_1^*=\int \left(\int \tau_2^0\,d\zeta\right)\,d\zeta,\quad
w_2^*= \int \left(\int w_2\,d\zeta\right)\,d\zeta$$
satisfy the  equation  (\ref{basiceq1Z}). Normalizing $w_1^*$
to get the asymptotic (\ref{Z}) we obtain
$$Z_{10}=\arctan \re^{s\zeta},$$
$Z_{10}\to \re^{s\zeta}$ as $\zeta\to\infty$.
The solution $Z_{20}$ asymptotic to $\re^{-s\zeta}$ as $\zeta\to\infty$
is got by normalizing $w^*_2$. The solutions
$Z_{30}=1$ and $Z_{40}=\zeta$  complete the basis in the space of solutions
of the equation  (\ref{basiceq1Z}).

The equations (\ref{basiceq2Y}), (\ref{basiceq3Y}) have the form
${\cal L}Y=W$. Integrating the last equation twice one gets
\begin{eqnarray}
\label{inh1}
{\cal L}_0 Y=\hat{W},\quad \hat{W}=\int\left(\int W\,d\zeta\right)\,d\zeta.
\end{eqnarray}
The particular solution of (\ref{inh1}) is given by
\begin{eqnarray}
\label{part1}
Y=-Y_{10}\int\hat{W}Y_{20}\,d\zeta+Y_{20}\int\hat{W}Y_{10}\,d\zeta.
\end{eqnarray}

The symbolic language {\tt Mathematica 4.0} was employed for
 construction of the  solutions
 of the form (\ref{part1})
 in  the explicit form. To get the asymptotics (\ref{Y}) we modify the
solutions of (\ref{basiceq2Y}), (\ref{basiceq3Y}) in the form (\ref{part1})
by adding a linear combination of the solutions
$Y_{\alpha 0}$
of the homogeneous equation
(\ref{basiceq1Y}).
 For example,
\begin{eqnarray*}
& & Y_{11}=\frac{c {\rm e}^{s\zeta}( {\rm e}^{2s\zeta}s\zeta-s\zeta-2)}
{s^3(1+ {\rm e}^{2s\zeta})^2},\quad Y_{11}\to
\frac{c}{1-c^2}\zeta{\re}^{-s\zeta},\quad\zeta\to\infty.
\end{eqnarray*}
To get $Y_{11}$ we add  $\gamma Y_{10}$
to the solution of (\ref{basiceq2Y})
in the form (\ref{part1}),
where the constant $\gamma$ is determined by (\ref{Y}).

The equations (\ref{basiceq2Z}), (\ref{basiceq3Z}) have the form
${\cal L}^*Z=\hat{W}$. Putting
$\phi=Z^{''}$ we get the equation of the form (\ref{inh1}) for the function
$\phi$. Using the formula (\ref{part1}) to obtain the particular solution, we
come back to $Z$ integrating $\phi$ twice.
Again, to get the asymptotics (\ref{Y}) we modify the
solutions of (\ref{basiceq2Z}), (\ref{basiceq3Z})
by adding a linear combination of the solutions
$Z_{\alpha 0}$
of the homogeneous equation.
 For example,
\begin{eqnarray*}
& &\hskip-0.7cmZ_{11}={-}\frac{c\arctan (\re^{s\zeta})}{s^3}{-}
\frac{c\zeta\arctan (\re^{s\zeta})}{s^2}{+}
\frac
{\ri c\,{\rm PolyLog}\,[2,-\ri\re^{s\zeta}]}{2s^3}{-}
\frac{\ri c\,{\rm PolyLog}\,[2,\ri\re^{s\zeta}]}{2s^3},
\\
& &\hskip-0.7cm  Z_{31}=\frac{\zeta}{1-c},\quad
Z_{32}=\frac{s^2(1+c)\zeta^2-
4(1+c)\,{\rm ln}\,(1+ {\rm e}^{2s\zeta})
}
{s^4(1-c)},\\
& &\hskip-0.7cmZ_{11}\to-\frac{c}{1-c^2}\zeta\re^{s\zeta},\quad Z_{32}\to \frac{\zeta^2}{(1-c)^2},\quad\zeta\to -\infty
\end{eqnarray*}
where ${\rm PolyLog}\,[2,y]=\int_y^0t^{-1}{\rm ln}(1-t)\,dt$, so that
\begin{eqnarray*}
& &\ri\,{\rm PolyLog}\,[2,-\ri\re^{s\zeta}]-
\ri\,{\rm PolyLog}\,[2,\ri\re^{s\zeta}]\to 2\re^{s\zeta},\quad
\zeta\to-\infty,\\
& &\ri\,{\rm PolyLog}\,[2,-\ri\re^{s\zeta}]-
\ri\,{\rm PolyLog}\,[2,\ri\re^{s\zeta}]\to \pi s \zeta+
2\re^{-s\zeta},\quad
\zeta\to\infty.
\end{eqnarray*}
To get $Z_{11}$ and $Z_{32}$ we modify the solution
of (\ref{basiceq2Z}), (\ref{basiceq3Z}),
obtained with the help of  (\ref{part1})
and further integration, by adding $\gamma_1 Z_{10}$ and $\gamma_2 Z_{40}$,
respectively, for the constants $\gamma_{1,2}$ to be determined from
(\ref{Z}).

The expressions for the other  solutions of
(\ref{basiceq1Y})-(\ref{basiceq3Z})   are rather long  and they are given in
Appendix A.

The coefficients at the powers of $\lambda$ for the expansion of the vector
functions ${\bf y}_{1,3}(\lambda,\zeta)$, ${\bf z}_{1,3}(\lambda,\zeta)$
are uniquely determined by the corresponding functions $Y_{ik}$, $Z_{ik}$,
$i=1,3$, $k=0,1,2$ (see the expressions at
the bottom of  (\ref{matr1}) and (\ref{matr2})).

To get the coefficients of the expansion series
in a zero  neighbourhood
for $\hat{D}(\lambda)$  the formula  (\ref{Evans}) was used.
We compute
\begin{eqnarray*}
\hat{D}(\lambda)=\det\left(
\begin{array}{cc}
a_{11}+b_{11}\lambda+d_{11}\lambda^2+O(\lambda^3)&a_{12}+b_{12}\lambda+
d_{12}\lambda^2+O(\lambda^3)\\
a_{21}+b_{21}\lambda+d_{21}\lambda^2+O(\lambda^3)&
a_{22}+b_{22}\lambda+d_{22}\lambda^2+O(\lambda^3)
\end{array}
\right),
\end{eqnarray*}
where
\begin{eqnarray*}
& &a_{mn}=0,\quad m,n=1,2, \quad b_{11}=b_{21}=d_{21}=0,\quad
b_{12}=\frac{\pi s^2}{2(1+c)},\quad  b_{22}=2,\\
& & d_{11}=\frac{1-2c^2}{2s^3},\quad d_{12}=-\pi\frac{2+3 c}{2(1+c)s},\quad
d_{22}=-\frac{4(1+c)}{(1-c)s}.
\end{eqnarray*}
Since   $a_{mn}=0$ the first nonzero coefficient of the expansion of
$\hat{D}(\lambda)$ is that one at the third power of $\lambda$ and, moreover,
this coefficient is completely determined by the coefficients at the
first and
second powers of $\lambda$  of the elements of the determinant.
Finally, we have
$$\hat{D}(\lambda)=\frac{1-2c^2}{s^3}\lambda^3+O(\lambda^4)
$$
in some neighbourhood  $U_0$ of the origin in the $\lambda$-plane.

According to Theorem 4.2
the normalized Evans function $D(\lambda)$, $D(\lambda)\to 1$ as
$|\lambda|\to\infty$, is given by
$$
D(\lambda)=\frac{\hat{D}(\lambda)}{
{\bf l}^\wedge(\lambda)\cdot {\bf r}^\wedge(\lambda)}.
$$
In a zero neighborhood one has
$$
{\bf l}^{\wedge}\cdot {\bf r}^{\wedge}=
\det\left(
\begin{array}{cc}
{\bf l}_1(\lambda)\cdot {\bf r}_1(\lambda)&{\bf l}_1(\lambda)\cdot {\bf
r}_3(\lambda)\\
{\bf l}_3(\lambda)\cdot {\bf r}_1(\lambda)&{\bf l}_3(\lambda)\cdot
{\bf r}_3(\lambda)
\end{array}
\right)
=-4(1-c^2)^{3/2}\lambda+O(\lambda^2).
$$
Hence, we get (\ref{DEvans}) and Theorem 1.2 is proved.

The computation of the normalized Evans function $D(\lambda)$ can be also
done by means of the integral formula (Pego and Weinstein [20], Theorem 1.11),
which in our case has the form
\begin{eqnarray}
\label{integr}
\dot{D}(\lambda)=- \frac{1}{{\bf l}^{\wedge}\cdot {\bf r}^{\wedge}}
\int\limits_{-\infty}^{\infty} {\bf z}^{\wedge}(\lambda,\zeta)\cdot
\dot{\cal M}^{\wedge}(\lambda,\zeta)\cdot
{\bf y}^{\wedge}(\lambda,\zeta)\,d\zeta,
\end{eqnarray}
where dot denotes differentiation with respect to $\lambda$.
The straightforward
computation gives
$$\dot{D}(\lambda)=-2\frac{1-2c^2}{4s^6}\lambda+O(\lambda^2),
$$
that is in full agreement with (\ref{DEvans}).
We list the expresion for the
coefficient at $\lambda^2$ of the
integrand in (\ref{integr}) in Appendix A.
\section{Conclusion and Discussion}
In this paper we  established
that for $c^2<1/2$ the leading coefficient of the Taylor
series of the normalized Evans
function  $D(\lambda)$
in a neighbourhood of zero is negative and thus,
$D(\lambda)$ has
to vanish somewhere  on the positive real axis. It follows then
from Theorem 4.1,
 that there exist the unstable eigenvalue
and associated with it unstable eigenfunction given by   (\ref{vid}).
This in its turn,  implies the exponential instability of the loop solitary
wave in the prescribed  velocity range.

In all cases known to us,   when the equation under analysis has a
hamiltonian structure and an invariant momentum associated with translational
invariance of the problem, there is the link between the convexity of the
momentum and the sign of the leading Taylor coefficient (Pego and Weinstein
[20], Alexander and Sachs [2], Bridges and Derks [7]). In  these
cases the existence of  the unstable  eigenfunction   is caused by the
translational symmetry. In the case under consideration the
momentum associated with
translational
invariance
${\cal T}(\omega) {\bf w}(\zeta)=\exp (\omega\partial_\zeta)\,{\bf w}(\zeta)=
{\bf w}(\zeta+\omega)$, $\omega\in {\Bbb R}$,
is $Q=Q({\bf w})$ (see Introduction).  One has
$$\frac{d Q(\phi_c)}{dc}=-\frac{8}{s^3}<0,
$$
i.\,e. the momentum is concave for all $c\in[0,1)$. Yet, the lead
coefficient  in the expansion of $D(\lambda)$ changes its sign at
$c=1/\sqrt{2}$ and, therefore, the above rule about the link is violated.
The reason of this violation is that the other symmetry than the translational
one is responsible for the existence of the unstable eigenfunction in our
case, namely the rotational symmetry around the $x_1$ axis:
${\cal G}(\varphi){\bf w}=\exp ({\cal A}\varphi)\,{\bf w}$,
$\varphi\in {\Bbb S}^1$,
where ${\cal A}$
is the block diagonal $6\times 6$ matrix with the blocks
$$
\left(
\begin{array}{ccc}
0&0&0\\0&0&1\\0&-1&0
\end{array}
\right).
$$

The question about stability of the loop solitary wave for
$c\in [1/\sqrt{2},1)$
 is still open, however
it looks plausible  that the impulse of the moving loop for
this range of speeds
is large enough to stabilize it and the loop manifests at least marginal
stability.
This  conjecture is based on the  analogy with a different  problem
of stability of solitary impulses in  a composite medium.
The linearized problem
about the solitary impulse in the composite is exactly the same as
the linear problem (2.2) of the paper, and it is known from the rigorous
nonlinear stability analysis
(and also from numerical evidence) that solitary impulses in question are
orbitally stable. It may mean that there is no unstable eigenfunction for
(2.2) for
$c>1/\sqrt{2}$.

Let us explain this analogy in more detail. The system of governing equations,
describing propagation  of quasi-transverse elastic waves
in some model of composite with nonlinear anisotropic elastic matrix
has the form (see e.g. Il'ichev [14])
\begin{eqnarray}
\label{basi}
\frac{\partial u_i}{\partial t_*}-\frac{\partial v_i}{\partial x}=0,
\hskip0.2cm
\rho_0 \frac{\partial v_i}{\partial t_*}-\frac{\partial}{\partial x}
\left(\frac{\partial{\cal\Phi}}{\partial u_i}\right)+m\frac{\partial^3 u_i}
{\partial x^3}=0,
\hskip0.2cm i=1,2.
\end{eqnarray}
Here
 $\rho_0$ is the average density,
$u_i$, $v_i$ --
gradients of displacements and particle velocities
in the wave front, respectively,
$\Phi$ is the elastic
potential, given by the expression
\begin{eqnarray*}
\Phi=\frac{1}{2} f(u_1^2+u_2^2)+\frac{1}{2}g(u_2^2-u_1^2)-\frac{1}{4}
\kappa (u_1^2+u_2^2)^2.
\end{eqnarray*}
The constants $f>0$,
 $g>0$, $\kappa$ and $m>0$ characterize
elastic modules,
anisotropy, nonlinearity and dispersion, respectively.
There is one-parametric anisotropic family of solitary wave solutions of
(\ref{basi}) for $\kappa>0$
\begin{eqnarray}
\label{asol}
u_1^0=\pm\sqrt{2\rho_0\kappa^{-1}(\mu_1-V^2)}{\rm sech }
\sqrt{\rho_0 m^{-1}(\mu_1-V^2)}(x-Vt_*),\hskip0.15cm u_2^0=0,
\end{eqnarray}
where $\mu_1=(f-g)/\rho_0$ and $V$ is the dimensional speed [14].
The solitary wave orbit is stable
 for $\mu_1/2\leq V^2<\mu_1$ ([14], Theorems 4, 5).
The results of numerical calculations reported in Bakholdin {\it et al} [5]
confirm this theoretical result, and also give the evidence that
solitary wave family (\ref{asol}) is unstable for $0\leq V^2<\mu_1/2$.

The equations (\ref{basi}), being linearized about the solitary wave
(\ref{asol})
for small perturbations $ \delta u_1=u_1-u_1^0$,
are transformed into one linear equation
\begin{eqnarray}
\label{line}
\frac{\partial^2 \delta u_1 }{\partial t_*^2}=\mu_1
\frac{\partial^2\delta
u_1}{\partial x^2}-
3\frac{\kappa}{\rho_0}\bigl(u_1^{0}\bigr)^2
\frac{\partial^2\delta
u_1}{\partial x^2}-
\frac{m}{\rho_0}
\frac{\partial^4\delta
u_1}{\partial x^4}.
\end{eqnarray}
Next, define the dimensionless speed  $c\in [0,1)$
and variables $\xi$, $t$, $\tau$ by
\begin{eqnarray*}
c=\frac{V}{\sqrt{\mu_1}},\quad
x=\sqrt{\frac{m}{\mu_1\rho_0}}\xi,\quad
t^*=\sqrt{\frac{m}{\rho_0\mu_1^2}}t,\quad
\delta u_1=\sqrt{\frac{\mu_1\rho_0}{\kappa}}\tau.
\end{eqnarray*}
Now, in the dimensionless variables, we have that solitary pulses in the
composite are
nonlinearly orbitally
stable
for $c^2>1/2$ (Theorems 4,5 of [14]).
In these  variables the linearized problem
(\ref{line}) has exactly the same form as the basic linear
problem for the loop stability (\ref{vozm})
with $p^0$ given by (\ref{sol}).
This result strongly prompts
that the linearized problem
(\ref{vozm}) can have no unstable eigenfunctions
for $c^2>1/2$, i.\,e. that
our loop solitary wave is at least marginally  stable inside this
range of speeds.
\begin{acknowledgements}
The present work is fulfilled under the financial support of the Russian
Foundation  for Basic Research grant No. 02-01-00729 and the President
Program of Support of Leading Scientific Schools grant No.
1697.2003.1.
\end{acknowledgements}
\appendix{ {\tt Mathematica} expressions for
$Y_{31}$,
$Y_{12}$, $Y_{32}$, $Z_{12}$
 and for the integrand in $(5.12)$}
In the  expressions below the notation $x$ instead of $\zeta$ is used.
\vskip0.2cm
\noindent
\begin{eqnarray*}
\Mvariable{Y_{31}}=
-\big({e^{-s\ x}}\ \big(32\ c\ {e^{s\ x}}+32\ {c^2}\ {e^{s\ x}}-96\ c\ {e^{3\ s\ x}}-96\ {c^2}\ {e^{3\ s\ x}}+2\ {e^{s\ x}}\ {s^2}-\\
2\ {e^{5\ s\ x}}\ {s^2}{+}
\pi \ {s^2}-{e^{2\ s\ x}}\ \pi \ {s^2}-17\ {e^{4\ s\ x}}\ \pi \ {s^2}+{e^{6\ s\ x}}\ \pi \ {s^2}{+}8\ c\ {e^{s\ x}}\ s\ x{+}  \\
 8\ {c^2}\ {e^{s\ x}}\ s\ x-48\ c\ {e^{3\ s\ x}}\ s\ x-48\ {c^2}\ {e^{3\ s\ x}}\ s\ x+8\ c\ {e^{5\ s\ x}}\ s\ x+  \\
 8\ {c^2}\ {e^{5\ s\ x}}\ s\ x+4\ {e^{s\ x}}\ {s^3}\ x-24\ {e^{3\ s\ x}}\ {s^3}\ x+4\ {e^{5\ s\ x}}\ {s^3}\ x-  \\
 12\ {e^{2\ s\ x}}\ \pi \ {s^3}\ x+12\ {e^{4\ s\ x}}\ \pi \ {s^3}\ x-2\ {s^2}\ \arctan [{e^{s\ x}}]+  \\
 18\ {e^{2\ s\ x}}\ {s^2}\ \arctan [{e^{s\ x}}]+18\ {e^{4\ s\ x}}\ {s^2}\ \arctan [{e^{s\ x}}]-  \\
 2\ {e^{6\ s\ x}}\ {s^2}\ \arctan [{e^{s\ x}}]+12\ i\ {e^{2\ s\ x}}\ {s^2}\ \Mfunction{PolyLog}[2,-i\ {e^{s\ x}}]-  \\
 12\ i\ {e^{4\ s\ x}}\ {s^2}\ \Mfunction{PolyLog}[2,-i\ {e^{s\ x}}]-12\ i\ {e^{2\ s\ x}}\ {s^2}\ \Mfunction{PolyLog}[2,i\ {e^{s\ x}}]+  \\
 12\ i\ {e^{4\ s\ x}}\ {s^2}\ \Mfunction{PolyLog}[2,i\ {e^{s\ x}}]\big)\big)\big/  \\
 \big(4\ (1+c)\ {{\big(1+{e^{2\ s\ x}}\big)}^2}\ {s^3}\big);
\end{eqnarray*}
\begin{eqnarray*}
\Mvariable{Y_{12}}=\frac{1}{2\ {{(1+{e^{2\ s\ x}})}^2}\ {s^6}}\big({e^{-s\ x}}\ \big({c^2}-17\ {c^2}\ {e^{2\ s\ x}}-{s^2}+ 13\ {e^{2\ s\ x}}\ {s^2}- \\
  4\ {e^{4\ s\ x}}\ {s^2}+
2\ {e^{s\ x}}\ \pi \ {s^2}-12\ {e^{3\ s\ x}}\ \pi \ {s^2}+2\ {e^{5\ s\ x}}\ \pi \ {s^2}-  \\
  14\ {c^2}\ {e^{2\ s\ x}}\ s\ x+
6\ {c^2}\ {e^{4\ s\ x}}\ s\ x-2\ {e^{2\ s\ x}}\ {s^3}\ x+2\ {e^{4\ s\ x}}\ {s^3}\ x- \\
  2\ {c^2}\ {e^{2\ s\ x}}\ {s^2}\ {x^2}+
2\ {c^2}\ {e^{4\ s\ x}}\ {s^2}\ {x^2}-4\ {e^{s\ x}}\ {s^2}\ \arctan [{e^{s\ x}}]+  \\
  24\ {e^{3\ s\ x}}\ {s^2}\ \arctan [{e^{s\ x}}]-
4\ {e^{5\ s\ x}}\ {s^2}\ \arctan [{e^{s\ x}}]-  \\
  8\ {e^{2\ s\ x}}\ {s^2}\ \log \big[1+{e^{-2\ s\ x}}\big]+
8\ {e^{4\ s\ x}}\ {s^2}\ \log \big[1+{e^{-2\ s\ x}}\big]\big)\big);\\
\end{eqnarray*}
\begin{eqnarray*}
\Mvariable{Y_{32}}=-\frac{1}{2\ (1+c)\ {{(1+{e^{2\ s\ x}})}^2}\ {s^6}}\big({e^{-s\ x}}\ \big(-128\ {c^2}\ {e^{s\ x}}-128\ {c^3}\ {e^{s\ x}}+  \\
256\ {c^2}\ {e^{3\ s\ x}}+256\ {c^3}\ {e^{3\ s\ x}}-2\ {e^{s\ x}}\ {s^2}+6\ c\ {e^{s\ x}}\ {s^2}-52\ {e^{3\ s\ x}}\ {s^2}-  \\
124\ c\ {e^{3\ s\ x}}\ {s^2}+6\ {e^{5\ s\ x}}\ {s^2}+14\ c\ {e^{5\ s\ x}}\ {s^2}-2\ \pi \ {s^2}+c\ \pi \ {s^2}+  \\
2\ {e^{2\ s\ x}}\ \pi \ {s^2}-10\ c\ {e^{2\ s\ x}}\ \pi \ {s^2}+34\ {e^{4\ s\ x}}\ \pi \ {s^2}+62\ c\ {e^{4\ s\ x}}\ \pi \ {s^2}-  \\
2\ {e^{6\ s\ x}}\ \pi \ {s^2}-5\ c\ {e^{6\ s\ x}}\ \pi \ {s^2}-64\ {c^2}\ {e^{s\ x}}\ s\ x-64\ {c^3}\ {e^{s\ x}}\ s\ x+  \\
192\ {c^2}\ {e^{3\ s\ x}}\ s\ x+192\ {c^3}\ {e^{3\ s\ x}}\ s\ x-18\ {e^{s\ x}}\ {s^3}\ x-50\ c\ {e^{s\ x}}\ {s^3}\ x+  \\
96\ {e^{3\ s\ x}}\ {s^3}\ x+192\ c\ {e^{3\ s\ x}}\ {s^3}\ x-14\ {e^{5\ s\ x}}\ {s^3}\ x-14\ c\ {e^{5\ s\ x}}\ {s^3}\ x+  \\
c\ \pi \ {s^3}\ x+24\ {e^{2\ s\ x}}\ \pi \ {s^3}\ x+11\ c\ {e^{2\ s\ x}}\ \pi \ {s^3}\ x-24\ {e^{4\ s\ x}}\ \pi \ {s^3}\ x-  \\
53\ c\ {e^{4\ s\ x}}\ \pi \ {s^3}\ x+c\ {e^{6\ s\ x}}\ \pi \ {s^3}\ x-8\ {c^2}\ {e^{s\ x}}\ {s^2}\ {x^2}-8\ {c^3}\ {e^{s\ x}}\ {s^2}\ {x^2}+  \\
48\ {c^2}\ {e^{3\ s\ x}}\ {s^2}\ {x^2}+48\ {c^3}\ {e^{3\ s\ x}}\ {s^2}\ {x^2}-8\ {c^2}\ {e^{5\ s\ x}}\ {s^2}\ {x^2}-8\ {c^3}\ {e^{5\ s\ x}}\ {s^2}\ {x^2}-  \\
2\ {e^{s\ x}}\ {s^4}\ {x^2}-6\ c\ {e^{s\ x}}\ {s^4}\ {x^2}+12\ {e^{3\ s\ x}}\ {s^4}\ {x^2}+36\ c\ {e^{3\ s\ x}}\ {s^4}\ {x^2}-  \\
2\ {e^{5\ s\ x}}\ {s^4}\ {x^2}-6\ c\ {e^{5\ s\ x}}\ {s^4}\ {x^2}-12\ c\ {e^{2\ s\ x}}\ \pi \ {s^4}\ {x^2}+12\ c\ {e^{4\ s\ x}}\ \pi \ {s^4}\ {x^2}+  \\
4\ {s^2}\ \arctan [{e^{s\ x}}]-2\ c\ {s^2}\ \arctan [{e^{s\ x}}]-100\ {e^{2\ s\ x}}\ {s^2}\ \arctan [{e^{s\ x}}]+  \\
22\ c\ {e^{2\ s\ x}}\ {s^2}\ \arctan [{e^{s\ x}}]+28\ {e^{4\ s\ x}}\ {s^2}\ \arctan [{e^{s\ x}}]-  \\
94\ c\ {e^{4\ s\ x}}\ {s^2}\ \arctan [{e^{s\ x}}]+4\ {e^{6\ s\ x}}\ {s^2}\ \arctan [{e^{s\ x}}]+  \\
10\ c\ {e^{6\ s\ x}}\ {s^2}\ \arctan [{e^{s\ x}}]+2\ {s^3}\ x\ \arctan [{e^{s\ x}}]+  \\
2\ c\ {s^3}\ x\ \arctan [{e^{s\ x}}]-18\ {e^{2\ s\ x}}\ {s^3}\ x\ \arctan [{e^{s\ x}}]-  \\
18\ c\ {e^{2\ s\ x}}\ {s^3}\ x\ \arctan [{e^{s\ x}}]-18\ {e^{4\ s\ x}}\ {s^3}\ x\ \arctan [{e^{s\ x}}]-  \\
18\ c\ {e^{4\ s\ x}}\ {s^3}\ x\ \arctan [{e^{s\ x}}]+2\ {e^{6\ s\ x}}\ {s^3}\ x\ \arctan [{e^{s\ x}}]+  \\
2\ c\ {e^{6\ s\ x}}\ {s^3}\ x\ \arctan [{e^{s\ x}}]-16\ c\ {e^{s\ x}}\ {s^2}\ \log \big[1+{e^{-2\ s\ x}}\big]+  \\
96\ c\ {e^{3\ s\ x}}\ {s^2}\ \log \big[1+{e^{-2\ s\ x}}\big]-16\ c\ {e^{5\ s\ x}}\ {s^2}\ \log \big[1+{e^{-2\ s\ x}}\big]+  \\
8\ {e^{s\ x}}\ {s^2}\ \log \big[1+{e^{2\ s\ x}}\big]+8\ c\ {e^{s\ x}}\ {s^2}\ \log \big[1+{e^{2\ s\ x}}\big]-  \\
48\ {e^{3\ s\ x}}\ {s^2}\ \log \big[1+{e^{2\ s\ x}}\big]-48\ c\ {e^{3\ s\ x}}\ {s^2}\ \log \big[1+{e^{2\ s\ x}}\big]+  \\
8\ {e^{5\ s\ x}}\ {s^2}\ \log \big[1+{e^{2\ s\ x}}\big]+8\ c\ {e^{5\ s\ x}}\ {s^2}\ \log \big[1+{e^{2\ s\ x}}\big]-  \\
i\ {s^2}\ \Mfunction{PolyLog}[2,-i\ {e^{s\ x}}]-2\ i\ c\ {s^2}\ \Mfunction{PolyLog}[2,-i\ {e^{s\ x}}]-15\ i\ {e^{2\ s\ x}}\ {s^2}\   \\
\Mfunction{PolyLog}[2,-i\ {e^{s\ x}}]+6\ i\ c\ {e^{2\ s\ x}}\ {s^2}\ \Mfunction{PolyLog}[2,-i\ {e^{s\ x}}]+  \\
33\ i\ {e^{4\ s\ x}}\ {s^2}\ \Mfunction{PolyLog}[2,-i\ {e^{s\ x}}]+
54\ i\ c\ {e^{4\ s\ x}}\ {s^2}\ \Mfunction{PolyLog}[2,-i\ {e^{s\ x}}]-  \\
i\ {e^{6\ s\ x}}\ {s^2}\ \Mfunction{PolyLog}[2,-i\ {e^{s\ x}}]-2\ i\ c\ {e^{6\ s\ x}}\ {s^2}\ \Mfunction{PolyLog}[2,-i\ {e^{s\ x}}]-  \\
12\ i\ {e^{2\ s\ x}}\ {s^3}\ x\ \Mfunction{PolyLog}[2,-i\ {e^{s\ x}}]-
12\ i\ c\ {e^{2\ s\ x}}\ {s^3}\ x\ \Mfunction{PolyLog}[2,-i\ {e^{s\ x}}]+  \\
12\ i\ {e^{4\ s\ x}}\ {s^3}\ x\ \Mfunction{PolyLog}[2,-i\ {e^{s\ x}}]+  \\
12\ i\ c\ {e^{4\ s\ x}}\ {s^3}\ x\ \Mfunction{PolyLog}[2,-i\ {e^{s\ x}}]+i\ {s^2}\ \Mfunction{PolyLog}[2,i\ {e^{s\ x}}]+  \\
2\ i\ c\ {s^2}\ \Mfunction{PolyLog}[2,i\ {e^{s\ x}}]+15\ i\ {e^{2\ s\ x}}\ {s^2}\ \Mfunction{PolyLog}[2,i\ {e^{s\ x}}]-  \\
6\ i\ c\ {e^{2\ s\ x}}\ {s^2}\ \Mfunction{PolyLog}[2,i\ {e^{s\ x}}]-33\ i\ {e^{4\ s\ x}}\ {s^2}\ \Mfunction{PolyLog}[2,i\ {e^{s\ x}}]-  \\
54\ i\ c\ {e^{4\ s\ x}}\ {s^2}\ \Mfunction{PolyLog}[2,i\ {e^{s\ x}}]+i\ {e^{6\ s\ x}}\ {s^2}\ \Mfunction{PolyLog}[2,i\ {e^{s\ x}}]+  \\
2\ i\ c\ {e^{6\ s\ x}}\ {s^2}\ \Mfunction{PolyLog}[2,i\ {e^{s\ x}}]+12\ i\ {e^{2\ s\ x}}\ {s^3}\ x\ \Mfunction{PolyLog}[2,i\ {e^{s\ x}}]+  \\
12\ i\ c\ {e^{2\ s\ x}}\ {s^3}\ x\ \Mfunction{PolyLog}[2,i\ {e^{s\ x}}]- \\
12\ i\ {e^{4\ s\ x}}\ {s^3}\ x\ \Mfunction{PolyLog}[2,i\ {e^{s\ x}}]-12\ i\ c\ {e^{4\ s\ x}}\ {s^3}\   \\
x\ \Mfunction{PolyLog}[2,i\ {e^{s\ x}}]+24\ i\ {e^{2\ s\ x}}\ {s^2}\ \Mfunction{PolyLog}[3,-i\ {e^{s\ x}}]+  \\
48\ i\ c\ {e^{2\ s\ x}}\ {s^2}\ \Mfunction{PolyLog}[3,-i\ {e^{s\ x}}]-24\ i\ {e^{4\ s\ x}}\ {s^2}\   \\
\Mfunction{PolyLog}[3,-i\ {e^{s\ x}}]-48\ i\ c\ {e^{4\ s\ x}}\ {s^2}\ \Mfunction{PolyLog}[3,-i\ {e^{s\ x}}]-  \\
24\ i\ {e^{2\ s\ x}}\ {s^2}\ \Mfunction{PolyLog}[3,i\ {e^{s\ x}}]-  \\
48\ i\ c\ {e^{2\ s\ x}}\ {s^2}\ \Mfunction{PolyLog}[3,i\ {e^{s\ x}}]+24\ i\ {e^{4\ s\ x}}\ {s^2}\ \Mfunction{PolyLog}[3,i\ {e^{s\ x}}]+  \\
48\ i\ c\ {e^{4\ s\ x}}\ {s^2}\ \Mfunction{PolyLog}[3,i\ {e^{s\ x}}]\big)\big);\\
\end{eqnarray*}
\begin{eqnarray*}
\Mvariable{Z_{12}}=-\frac{1}{2\ {s^6}}\Big(-{c^2}\ {e^{s\ x}}+{e^{s\ x}}\ {s^2}+7\ {c^2}\ \arctan [{e^{s\ x}}]+  \\
13\ {s^2}\ \arctan [{e^{s\ x}}]+2\ {c^2}\ s\ x\ \arctan [{e^{s\ x}}]+2\ {s^3}\ x\ \arctan [{e^{s\ x}}]-  \\
2\ {c^2}\ {s^2}\ {x^2}\ \arctan [{e^{s\ x}}]-16\ {s^2}\ \arctan [{e^{s\ x}}]\ \log [2]-  \\
i\ {c^2}\ \Mfunction{PolyLog}[2,-i\ {e^{s\ x}}]-5\ i\ {s^2}\ \Mfunction{PolyLog}[2,-i\ {e^{s\ x}}]+  \\
2\ i\ {c^2}\ s\ x\ \Mfunction{PolyLog}[2,-i\ {e^{s\ x}}]+i\ {c^2}\ \Mfunction{PolyLog}[2,i\ {e^{s\ x}}]+  \\
5\ i\ {s^2}\ \Mfunction{PolyLog}[2,i\ {e^{s\ x}}]-2\ i\ {c^2}\ s\ x\ \Mfunction{PolyLog}[2,i\ {e^{s\ x}}]+  \\
8\ i\ {s^2}\ \Mfunction{PolyLog}\big[2,\frac{1}{2}-\frac{1}{2}\ i\ {e^{s\ x}}\big]-8\ i\ {s^2}\ \Mfunction{PolyLog}\big[2,\frac{1}{2}+\frac{1}{2}\ i\ {e^{s\ x}}\big]-  \\
2\ i\ {c^2}\ \Mfunction{PolyLog}[3,-i\ {e^{s\ x}}]-2\ i\ {s^2}\ \Mfunction{PolyLog}[3,-i\ {e^{s\ x}}]+  \\
2\ i\ {c^2}\ \Mfunction{PolyLog}[3,i\ {e^{s\ x}}]+2\ i\ {s^2}\ \Mfunction{PolyLog}[3,i\ {e^{s\ x}}]\Big);\\
\end{eqnarray*}
\begin{eqnarray*}
\Mvariable{\rm Integrand}=
\big({e^{s\ x}}\ \big(-16\ {c^3}\ {e^{s\ x}}+16\ {c^5}\ {e^{s\ x}}+16\ {c^3}\ {e^{3\ s\ x}}- \\
16\ {c^5}\ {e^{3\ s\ x}}+
16\ c\ {e^{s\ x}}\ {s^2}-
16\ {c^3}\ {e^{s\ x}}\ {s^2}-16\ c\ {e^{3\ s\ x}}\ {s^2}+  \\
16\ {c^3}\ {e^{3\ s\ x}}\ {s^2}+2\ c\ \pi \ {s^2}-
2\ {c^2}\ \pi \ {s^2}-
12\ c\ {e^{2\ s\ x}}\ \pi \ {s^2}+  \\
12\ {c^3}\ {e^{2\ s\ x}}\ \pi \ {s^2}+2\ c\ {e^{4\ s\ x}}\ \pi \ {s^2}+
2\ {c^2}\ {e^{4\ s\ x}}\ \pi \ {s^2}-
4\ {c^3}\ {e^{4\ s\ x}}\ \pi \ {s^2}+ \\
\pi \ {s^4}-c\ \pi \ {s^4}-
{e^{4\ s\ x}}\ \pi \ {s^4}+c\ {e^{4\ s\ x}}\ \pi \ {s^4}-
{c^2}\ \pi \ {s^3}\ x+
{c^3}\ \pi \ {s^3}\ x+  \\
6\ {c^2}\ {e^{2\ s\ x}}\ \pi \ {s^3}\ x-6\ {c^3}\ {e^{2\ s\ x}}\ \pi \ {s^3}\ x-
{c^2}\ {e^{4\ s\ x}}\ \pi \ {s^3}\ x+
{c^3}\ {e^{4\ s\ x}}\ \pi \ {s^3}\ x- \\
c\ \pi \ {s^5}\ x+6\ c\ {e^{2\ s\ x}}\ \pi \ {s^5}\ x-
c\ {e^{4\ s\ x}}\ \pi \ {s^5}\ x-
8\ c\ {s^2}\ \arctan [{e^{s\ x}}]- \\
8\ {c^2}\ {s^2}\ \arctan [{e^{s\ x}}]-
8\ {c^3}\ {s^2}\ \arctan [{e^{s\ x}}]+
48\ c\ {e^{2\ s\ x}}\ {s^2}\ \arctan [{e^{s\ x}}]+  \\
96\ {c^2}\ {e^{2\ s\ x}}\ {s^2}\ \arctan [{e^{s\ x}}]+
48\ {c^3}\ {e^{2\ s\ x}}\ {s^2}\ \arctan [{e^{s\ x}}]- \\
8\ c\ {e^{4\ s\ x}}\ {s^2}\ \arctan [{e^{s\ x}}]-
24\ {c^2}\ {e^{4\ s\ x}}\ {s^2}\ \arctan [{e^{s\ x}}]-  \\
8\ {c^3}\ {e^{4\ s\ x}}\ {s^2}\ \arctan [{e^{s\ x}}]-
4\ {s^4}\ \arctan [{e^{s\ x}}]+  \\
4\ {e^{4\ s\ x}}\ {s^4}\ \arctan [{e^{s\ x}}]+
2\ i\ {c^2}\ {s^2}\ \Mfunction{PolyLog}[2,-i\ {e^{s\ x}}]- \\
12\ i\ {c^2}\ {e^{2\ s\ x}}\ {s^2}\ \Mfunction{PolyLog}[2,-i\ {e^{s\ x}}]+
2\ i\ {c^2}\ {e^{4\ s\ x}}\ {s^2}\ \Mfunction{PolyLog}[2,-i\ {e^{s\ x}}]-  \\
2\ i\ {c^2}\ {s^2}\ \Mfunction{PolyLog}[2,i\ {e^{s\ x}}]+12\ i\ {c^2}\ {e^{2\ s\ x}}\ {s^2}\ \Mfunction{PolyLog}[2,i\ {e^{s\ x}}]-  \\
2\ i\ {c^2}\ {e^{4\ s\ x}}\ {s^2}\ \Mfunction{PolyLog}[2,i\ {e^{s\ x}}]\big)\big)\big/  \\
\big((-1+c)\ (1+c)\ {{\big(1+{e^{2\ s\ x}}\big)}^3}\ {s^2}\big);\\
\end{eqnarray*}
The rest of solutions  of (\ref{basiceq1Y})-(\ref{basiceq3Z})
with their asymptotics are given in sec.
5.
The listed  functions have the asymptotics
\begin{eqnarray*}
& &
Y_{12}\to\left(\frac{c^2x^2}{(1-c^2)^2}+\frac{1+2c^2}{(1-c^2)^{5/2}}x\right)
\re^{-sx},\\
& &
Y_{31}\to-\frac{x}{1-c},\quad Y_{32}\to\frac{x^2}{(1-c)^2},\,\,{\rm as}\,\,x\to\infty,
\end{eqnarray*}
and
\begin{eqnarray*}
Z_{12}\to
\left(\frac{c^2x^2}{(1-c^2)^2}-\frac{1+2c^2}{(1-c^2)^{5/2}}x\right)
\re^{sx},\,\,{\rm as}\,\,x\to-\infty.
\end{eqnarray*}
The function   $P(x)={\rm PolyLog}[3,\ri\re^{x}]$ above is defined
via
$$\frac{d P(x)}{dx}={\rm PolyLog}[2,\ri\re^x],$$
and has the asymptotic behaviour
$$\Imag P(x)\to - {\re^{x}},\,\, {\rm as}\,\, x\to-\infty;
\quad \Imag P(x)\to -\frac{\pi}{4}x^2,\,\,{\rm as}\,\, x\to\infty.
$$

\label{lastpage}

\end{document}